\documentclass[aps,prl,twocolumn,groupedaddress,showpacs]{revtex4}
\usepackage{graphicx,amsmath,amssymb,amsfonts,latexsym,color,dcolumn,bm}
\usepackage{verbatim}

\begin{document}

\title{Novel E-beam lithography technique for in-situ junction fabrication: the controlled undercut}

\author{F. Lecocq, C. Naud, I. M. Pop, Z. H. Peng, I. Matei,  T. Crozes, T. Fournier, W. Guichard and O. Buisson}

\address{Institut N\'eel, CNRS and Universit\'{e} Joseph Fourier,
38042 Grenoble, France}%

\begin{abstract}
We present a novel shadow evaporation technique for the realization of junctions and capacitors. The design by E-beam lithography of strongly asymmetric undercuts on a bilayer resist enables in-situ fabrication of junctions and capacitors without the use of the well-known suspended bridge\cite{Dolan}. The absence of bridges increases the mechanical robustness of the resist mask as well as the accessible range of the junction size, from $10^{-2}\mu m^2$ to more than $10^{4}\mu m^2$. We have fabricated Al/AlO$_x$/Al Josephson junctions, phase qubit and capacitors using a 100kV E-beam writer. Although this high voltage enables a precise control of the undercut, implementation using a conventional 20kV E-beam is also discussed. The phase qubit coherence times, extracted from spectroscopy resonance width, Rabi and Ramsey oscillations decay and energy relaxation measurements, are longer than the ones obtained in our previous samples realized by standard techniques. These results demonstrate the high quality of the junction obtained by the controlled undercut technique.

\end{abstract}



\maketitle

\section{Introduction}
	On-chip metallic junctions are the building blocks for a wide variety of nanoelectronic devices such as single electron transistors \cite{Coulombblockade}, spin-based electronic devices \cite{Yang, Valenzuela} and superconducting circuits such as SQUIDs \cite{SQUID}, voltage standard circuits\cite{VoltageStandard}, RSFQ logic circuits\cite{RSFQ}, nano-fridges\cite{Nanofridges} and superconducting qubits \cite{reviewqubit09}.
The Shadow Evaporation Technique (ShET)\cite{Dolan} with a suspended bridge of resist appears as a very useful technique to realize these circuits. For ShET the two metal evaporations, as well as the oxidation to obtain the tunnel barrier, are made without breaking the vacuum therefore enabling high quality junctions. ShET is simple (one lithography step) and flexible (independent on the metal choice) and enables submicron size tunnel junctions. These performances make this technique often preferable compared to the Multi-Layers Technique (MLT)\cite{Gurvitch,Dolata}. However ShET presents important limitations due to the suspended bridge. Overcoming these limitations would open the way for designing new types of circuits. One of the ShET disadvantage is the fragility of the bridge which makes any etching by plasma difficult in order to clean the substrate surface before evaporation. In addition, the bridge prevents from an efficient cleaning of the junction surface before evaporation since it is located just above the junction. As a consequence the resist residues can contaminate the tunnel barrier and alter its oxide quality. Moreover the mechanical strains on the suspended bridge make impossible the fabrication of large tunnel junctions (typically larger than 10$\mu m^2$) and large capacitors.
	
			In this paper we report on a Josephson junction phase qubit made by a novel technique using angle evaporations but without using suspended bridges. This technique, called hereafter Controlled Undercut Technique (CUT), is based on the control of strongly asymmetric undercuts in a bilayer resist. By adjusting the undercut position and its depth,
we select for each angle of evaporation whether the metal
will be deposited onto the substrate or on the resist wall
which will be removed after lift-off. By this way we can control which wire will be connected to the junction.
 Using the CUT, we have fabricated a Camelback phase qubit based on a zero-current bias SQUID. Josephson junction current-voltage (IV) and qubit coherence properties have been characterized and compared to our previous Camelback phase qubits  made by Multi-Layers Technique (MLT)\cite{Hoskinson09} and ShET techniques\cite{Fay_08}.

\section{Controlled undercut technique}

\begin{figure}[ht]
\begin{minipage}[t]{240pt}
\includegraphics[scale=0.4]{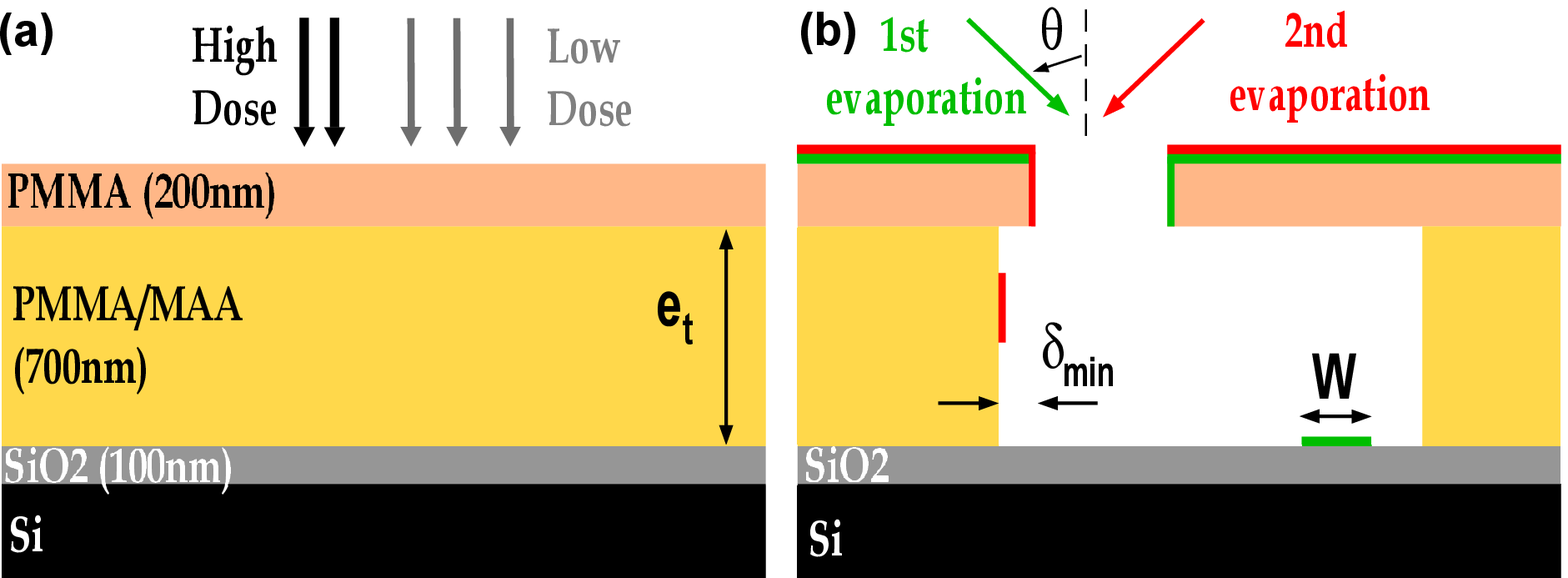}
\end{minipage}
\\
\begin{minipage}[b]{240pt}
\includegraphics[scale=0.8]{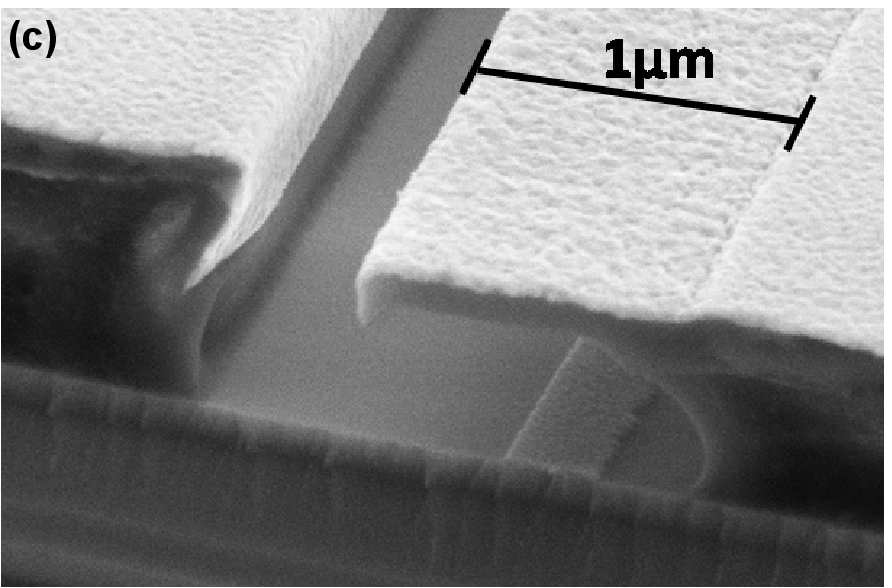}
\end{minipage}
\caption{ Schematic cross sectional diagram of the CUT. (a) During the E-beam exposure. (b) During the 
evaporation after the resist development.  (c) A cross section SEM image before lift-off obtained by cleaving the Si wafer.
The first evaporation wire is below the undercut. The second evaporation wire is on the resist wall. }
\label{fig1}
\end{figure}

	The CUT was developed using a 200nm thick PMMA imaging layer on top of a 700nm thick copolymer PMMA/MAA support layer spun on a $Si/SiO_2$ wafer. An E-beam writer operating at 100kV draws two successive types of patterns. A first one using a high dose exposure defines the opened wires in the imaging layer. The second pattern using a low dose defines the undercut in the support layer (Fig.\ref{fig1}(a)). These two different patterns are possible because the PMMA sensitivity is around three times lower than the PMMA/MAA one. After development (MIBK(1):IPA(3) during 30s and rinsing with pure IPA) we obtain a strongly asymmetric undercut (Fig.\ref{fig1}(b)). Typically we achieve on one side a designed undercut with a depth up to more than 1$\mu$m while on the other side we observe an undesired residual undercut smaller than 50nm. A light oxygen plasma RIE cleans the resist residues on the wafer. The next step consists of two evaporations with angle $\theta=-45^{\circ}$ and $\theta=+45^{\circ}$, separated by an in-situ oxidation (Fig.\ref{fig1}(b)). The first evaporation produces a wire on the substrate whereas the second one is deposited on the resist wall. After resist lift-off only one wire remains on the substrate. In the case of an undercut designed on the other side, the situation is symmetrically reversed: the second evaporation is deposited on the substrate and the first one on the resist and removed after lift-off. Therefore the CUT enables to design undercut patterns such that we select for each evaporation which connecting wire remains or not on the substrate.

\begin{figure}[ht]
\includegraphics[width=0.45\textwidth]{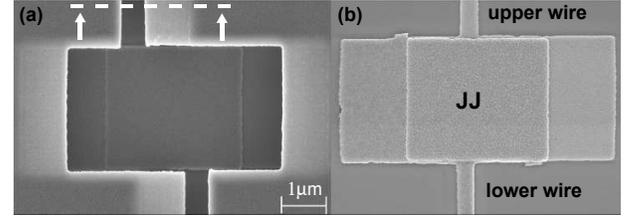}
\caption{(color online) (a) SEM pictures of a junction (central part) and its two connecting wires before lift-off. The bright areas indicate the undercuts, the dark area the developped resist. The white dashed line and two arrows indicate the Fig\ref{fig1}(c) cut.  (b) SEM pictures after lift-off. The upper and lower wires which correspond respectively to the first and the second evaporation, connect the bottom and top electrode of the Al junction to the circuit. The alignement of the two Al connecting wires from the unaligned resist mask is realized because of the two opposite evaporation angles.}   
\label{fig2}
\end{figure}

 Fig. \ref{fig2}(a) illustrates the typical resist mask design to realize a junction using the CUT. 
 The central opened area defines the junction. The two opened lines, on both side of the central area, define the two wires connecting the junction to the external circuit. A $1{\mu}m$ deep undercut is present on the right of the upper connecting wire and on the left of the lower connecting wire. A cross section of the upper wire corresponds to Fig.\ref{fig1}(b) and (c). To realize a junction we perform two evaporations with opposite angles as mentioned  previously. The first evaporation, coming from the left in Fig. \ref{fig2}(a), only deposits on the substrate the bottom electrode of the junction and the upper wire. The evaporation of the lower wire is deposited on the resist wall. Similarly, the second evaporation, coming from the right, deposits on the substrate the top electrode of the junction, the lower wire on the substrate and the upper wire on the resist wall. Fig.\ref{fig2}(b) shows the junction after lift-off. The two evaporations are shifted by about $1.6$ ${\mu}m$ because of the different angles.

\begin{figure}[ht]
\includegraphics[width=0.5\textwidth]{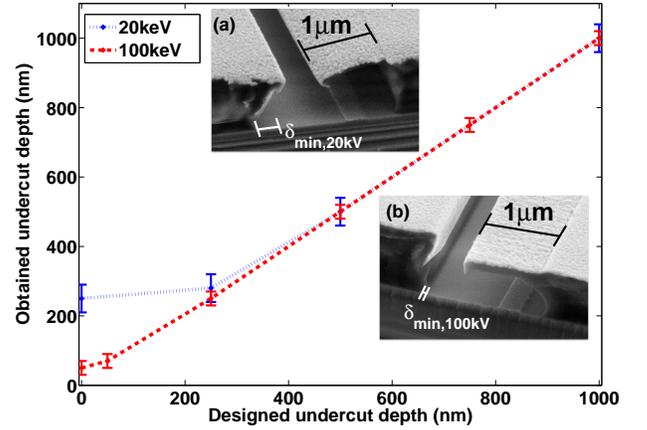}
\caption{ Obtained undercut depth after development versus designed undercut depth for electron beam energy of $20 keV$ and $100 keV$. Insets a) and b) present SEM images of an asymmetry undercut, with the minimum achievable undercut depth $\delta_{min}$ on one side and a $1{\mu}m$ deep undercut on the other side, for electron energy of $20 keV$ and $100 keV$\label{fig3}, respectively.}
\end{figure}

	The connecting wires must have a width $W$ which satisfy the condition $W < tg(\theta)*e_{t} -\delta_{min}$ where $e_{t}$ is the support layer thickness and $\delta_{min}$ the residual undercut depth (see Fig.\ref{fig1}). Indeed if the width is too large, both angle evaporations will be deposited onto the substrate and will short-circuit the junction. This width is the key parameter to quantify the feasibility of the CUT. It increases with the resist thickness and linearly decreases with $\delta_{min}$.
	
	In order to quantify the undercut control in our process, we have measured the obtained undercut depth after development as a function of the designed one (see Fig.\ref{fig3}). The data were obtained by measuring the undercut depth before lift-off on each side of a 250nm large wire. For deep designed undercuts (above 300nm), the measured undercut corresponds to the designed one and a good undercut control is reached independently of the E-beam voltage. For zero designed undercut, a residual undercut is observed which corresponds to $\delta_{min}$. It is mainly produced by proximity effect and is strongly dependent on the E-beam voltage \cite{handbook}. As shown in the insets in Fig.\ref{fig3}, $\delta_{min}$ drops from $250nm$ to $40nm$ when the voltage increases from 20kV to 100kV. This effect shows the advantage of high voltage for the control of strongly assymmetric undercuts.

	 Therefore we have developed the CUT using a 100kV E-beam to minimize $\delta_{min}$ and maximize $W$. Although preliminary lithography at 20kV was unsuccessful, increasing the  bottom resist thickness should make this new technique feasible even at this low voltage. We would also like to mention that a better undercut control can be achieved by using either another resist \cite{Aumentado} or another developper instead of MIBK \cite{Yasin,Stein}. This would minimize the residual development of unexposed resist.

\begin{figure}[ht]
\includegraphics[width=0.45\textwidth]{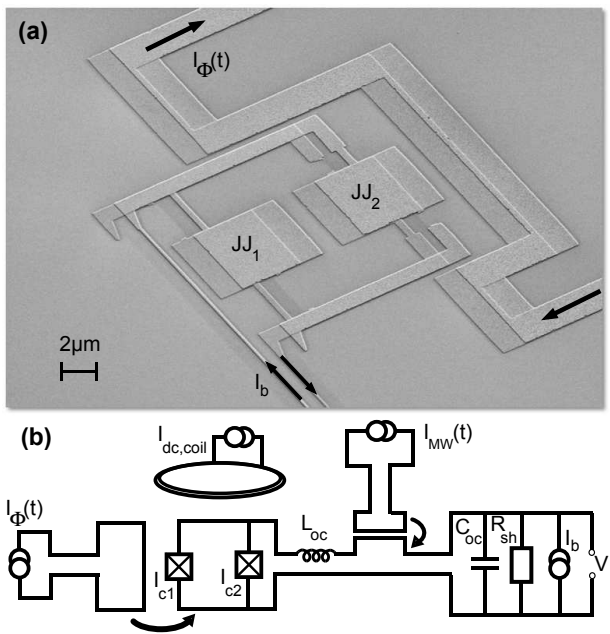}
\caption{(a) SEM image of a Camelback phase qubit based on a dc SQUID after lift-off.(b) The dc SQUID phase qubit is controlled by a dc bias current $I_b$ and flux $\Phi_b$, manipulated by a MW current and measured by a nanosecond flux pulse. The qubit is decoupled from the environment by a large kinetic inductance $L_e=10nF$ and capacitance $C_e=200pF$ and is shunted by a resistance $R_{sh}=20k\Omega$.
\label{fig4}}
\end{figure}

\section{Camelback phase qubit realization}

A Camelback phase qubit has been fabricated using the CUT. Fig.\ref{fig4}.a shows a SEM image of the device after lift-off. The SQUID loop with the two 10$\mu m^{2}$-JJ (central part) is connected by two $100$ nm-thin and $200$ $\mu m$-long current bias lines (visible on the left bottom side). The larger wire on the right side is the antenna for the high frequency bias flux. Fig.\ref{fig4}.b shows the electrical circuit for the measurement of the Camelback phase qubit.

The Al/AlO$_x$/Al qubit tunnel junctions and the device were formed by evaporating from a $-45^{\circ}$ angle $15$ nm of Al in a high vaccum chamber, oxidizing for $10$ min in $30$ mbar of oxygen, depositing at a $45^{\circ}$ angle $30$ nm of Al, and then lifting off the pattern. The IV characteristics at low temperature of the dc SQUID gives the normal resistance of the two parallel tunnel junctions, R$_n$=246 $\Omega$. The total  critical current, $I_{c,1}+I_{c,2}= 1.426 \mu A$, measured at zero magnetic flux, is consistent with the Ambegaokar-Baratoff formula, $\frac{\pi \Delta} {2eR_n}$=$1.405 \mu A$ where $\Delta=220\mu eV$ is the Al superconducting gap extracted from IV characteristics. The subgap resistance is larger than the 20 $k\Omega$ of the shunt resistance and the retrapping current is less than 7 $nA$. Moreover the measurement of the $I_c$ versus flux shows a very small critical current asymmetry between the two JJ of the SQUID, $I_{c,1}-I_{c,2}=3nA$. The small values of the retrapping current and critical current asymmetry as well as the large subgap resistance is a first proof of the high quality of the JJ made by the CUT.
 
\begin{figure}[ht]
\includegraphics[width=0.5\textwidth]{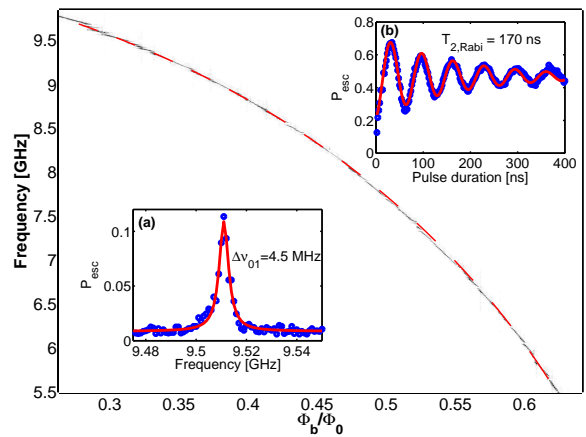}
\caption{Camelback phase qubit dynamics. (a) Spectroscopy:
$P_{esc}$ versus reduced flux and microwave frequency for a current bias $I_b=2nA$.
$P_{esc}$ is enhanced when the frequency matches $\nu_{01}$.
Dark and bright gray scale correspond to high and small $P_{esc}$.
Dashed line is $\nu_{01}$ versus reduced flux derived from the theory presented in Ref.\cite{Hoskinson09}. 
Insets (a): Qubit spectroscopy at $\Phi/\Phi_0=0.308$ and $I_b=2nA$. In this particular frequency window TLS are not visible. The fit is made by lorentzian curve.
Inset (b): Rabi oscillation at the same working point.
\label{fig5}}
\end{figure}

We present results on the quantum dynamics at zero current bias when the potential is quartic and the qubit is insensitive to current noise\cite{Hoskinson09}. Spectroscopy measurements were performed by applying a current-bias microwave pulse of duration $800ns$. An adjusted nanosecond flux pulse is applied just after the microwave pulse and produces an escape whose probability $P_{esc}$ is proportional to the occupancy of the excited qubit state\cite{Claudon_PRB2007}. Spectroscopy with a resonant peak corresponding to the transition between level $|0\rangle$ and level  $|1\rangle$ of the qubit is shown in Fig.\ref{fig5} inset ($a$).
The phase qubit presents resonance peaks at $\nu_{01}$ with a width of $\Delta\nu_{0,1}=4 MHz$.
In Fig.\ref{fig5}, escape probability is plotted versus bias flux $\Phi_b$ and microwave frequency. The experimental energy spectrum $\nu_{01}(\Phi_b)$ is precisely described by the Camelback potential theory\cite{Hoskinson09}.
Relaxation time of about $200 ns$ has been measured on the qubit. Rabi oscillations were observed with a $170 ns$ exponential decay time (Fig.\ref{fig5} inset ($b$)). From Ramsey oscillations  we deduce a coherence time of about $100$ ns which is consistent with the time extracted from the resonance peak width. This coherence time is about five times longer than the one measured in our previous samples made by SMT\cite{Hoskinson09} or ShET \cite{Fay_08}. These qubit coherence properties measurements validate our novel technique for the fabrication of tunnel junctions. We have also successfully fabricated in-situ large capacitances up to $10^{4}\mu m^2$.

\section{Discussion}
 
The spectroscopy also probes the microscopic Two-Level Systems (TLS) which are coupled to the qubit\cite{Cooper_PRL04,Hoskinson09,Lisenfeld_PRB10,Palomaki_PRB10}. A density of about $25$
visible TLS per GHz is measured with a typical coupling strenght ranging between $10MHz$ and $40MHz$. These values are similar to the ones obtained on phase qubits with comparable JJ size realized by MLT and ShET techniques\cite{Cooper_PRL04,Lisenfeld_PRB10,Palomaki_PRB10}. The CUT does not reduce the TLS density. However as it has been demonstrated\cite{Steffen_PRL06}, this limitation can be avoided by reducing the JJ size.

  	The CUT has specific advantages compared to the standard techniques. It preserves the advantages of the ShET compared to the MLT with a single lithography step and submicron junction area. It enables realization of very large junctions as it eliminates the mechanical constraints imposed by the suspended bridges. It also exhibits higher robustness and reproducibility. The stronger mechanical strenght as well as the absence of a bridge located just above the junction area enables direct and efficient cleaning of the junction area by reactive ion etching or ion milling. All these
 	improvements lead to a better junction quality. 
 In comparison with the recent technique	based on deep resist trenches\cite{Potts,Gladchenko_NatPhys09}, CUT also presents significant advantages. Indeed since the deep trenches technique needs two perpendicular evaporations, it requires a more complex circuit pattern compared to circuits made by ShET and CUT. More sophisticated circuits can be reached with CUT by using for example three different evaporation angles and
two different oxidations. However the CUT, like all other lift-off techniques needs particular attention and developments when refractory materials are deposited\cite{Dubos}. We would also like to point out that tunnel junctions using the CUT and suspended bridge technique can be realized at the same time.

\section{Conclusion}

In conclusion, we present an original lithography process for in-situ junctions fabrication. This technique is based on the complete control of the bilayer resist undercut. The CUT reduces the usual mechanical limitations inherent to other suspended shadow-mask techniques and enables an improvement of the junction quality. This novel method is able to realize junctions and on-chip capacitors with an extended size range, from $10^{-2}\mu m^2$ to more than $10^{4}\mu m^2$. The IV characteristics as well as coherence properties of the Camelback phase qubit demonstrate the high quality of the junction. These results definitely validate the CUT as a promissing method to realize junctions and capacitors for a wide range of devices.

\vspace{0.1cm}
\textbf{Acknowledgments}
\vspace{0.1cm}
 
We thank S. Decossas, H. Haas, T. Meunier, J.L. Thomassin and L. Vila for their support on the E-beam writer. We are grateful to B. Fernandez and C. Lemonias of the Nanofab facility in Grenoble for fruitful discussions and technical support. We also acknowledge the technical support of the PTA facility in CEA Grenoble.
This work was supported by the european EuroSQIP and SOLID projects and by the french ANR \textquotedblright QUANTJO\textquotedblright.

\end{document}